\documentclass[final]{ustcstep}
\usepackage{multirow,color}
\usepackage{bbding}
\usepackage{amssymb}
\usepackage{ulem}
\usepackage{color}
\usepackage[dvips]{graphicx}
\usepackage[square]{natbib}
\usepackage{abstract}

\def\gsim{\;\lower4pt\hbox{${\buildrel\displaystyle >\over\sim}$}\;}
\def\lsim{\;\lower4pt\hbox{${\buildrel\displaystyle <\over\sim}$}\;}
\def\grls{\;\lower4pt\hbox{${\buildrel\displaystyle >\over <}$}\;}

\newcommand\addr[2]{{\footnotesize \it $^{#1}$#2}\\}

\begin{document}

\title{When and how does a prominence-like jet gain kinetic energy?}

\author{Jiajia Liu,$^{1}$ Yuming Wang,$^{1,*}$ Rui Liu,$^1$ Quanhao Zhang,$^{1}$ Kai Liu,$^{1}$ Chenglong Shen,$^1$ and S. Wang$^1$ \\[1pt]
\addr{1}{CAS Key Laboratory of Geospace Environment,
Department of Geophysics and Planetary Sciences,University of Science \& Technology}
\addr{ }{ of China, Hefei, Anhui 230026, China}
\addr{*}{Correspondence and requests for materials should be addressed to Yuming Wang (ymwang@ustc.edu.cn)}}

\maketitle
\tableofcontents

\begin{abstract}
Jet, a considerable amount of plasma being ejected from chromosphere
or lower corona into higher corona, is a common phenomenon. Usually
a jet is triggered by a brightening or a flare, which provides the
first driving force to push plasma upward. In this process, magnetic
reconnection is thought to be the mechanism to convert magnetic
energy into thermal, non-thermal and kinetic energies. However, most
jets could reach an unusual high altitude and end much later than
the end of its associated flare. This fact implies that there is
another way to continuously transfer magnetic energy into kinetic
energy even after the reconnection. The whole picture described
above is well known in the community, but how and how much magnetic
energy is released through the way other than the reconnection is
still unclear. Here, through studying a prominence-like jet observed
by SDO/AIA and STEREO-A/EUVI, we find that the continuous relaxation
of the post-reconnection magnetic field structure is an important
process for a jet to climb up higher than it could through only
reconnection. The kinetic energy of the jet gained through the
relaxation is 1.6 times of that gained from the reconnection. The
resultant energy flux is hundreds of times larger than the flux
required for the local coronal heating, suggesting that such jets
are a possible source to keep corona hot. Furthermore, rotational
motions appear all the time during the jet. Our analysis suggests
that torsional Alfv\'en waves induced during reconnection could not
be the only mechanism to release magnetic energy and drive jets.
\end{abstract}

\section{Introduction}
Solar jets are a ubiquitous activity in the solar atmosphere, from
active regions, quiet Sun region to polar region. According to their
size and observed wavelengths, jets could be classified as surge
\citep[e.g.,][]{Newton1934, Rust1968, Roy1973, Xu1984, Canfield1996,
Jibben2004}, multi-wavelength (UV, EUV to X-ray) jets
\citep[e.g.,][]{Schmieder1988, Shibata1992, Cirtain2007,
Culhane2007, Liu2009, ShenY2011, Tian2012} and spicules
\citep[e.g.,][]{DePontieu2007b, DePontieu2007,
Shibata2007}. These jets carry lots of mass and energy
from low solar atmosphere into corona, and therefore are thought to
play an important role in coronal heating and solar wind
acceleration \citep[e.g.,][]{Shibata1996, Shibata2007,
Tsiropoula2004, DePontieu2007}.

Previous studies have shown that the length of solar jets range from
about one to several hundreds megameter, the speed could be from
10 to thousand kilometers per second, and the lifetime spreads from
minutes to hours \citep[e.g.,][]{Shibata1996,
Cirtain2007, DePontieu2007b}.
Usually, a jet has two components: a hot component
and a cool component, which are mainly distributed in the temperature
of soft X-ray and 304\AA, respectively \citep[e.g.,][]{Moore2013}.
Either hot or cool component could be dominant. Therefore some
jets are visible in H$\alpha$ or 304\AA\ passbands, while some jets
are visible in EUV or X-ray observations \citep[e.g.,][]{ShenY2011, Srivastava2011}.
Although different types of jets have different properties, some common
phenomena could be found in most cases. The first common phenomenon
is flaring, a manifestation of magnetic field reconnection. Except
type I spicules \citep{DePontieu2007b}, stronger or
weaker flaring could be always found at the jet root. It is believed
to be the initial and major driver of a jet. However, observations
showed that the initial speed of a jet usually is too small to make
it to a height as observed  \citep[e.g.,][]{Roy1973, Liu2009, ShenY2011}, suggesting that some
additional force after the flaring must act on the jet plasma.

\begin{figure*}
\begin{center}
\includegraphics[width=\hsize]{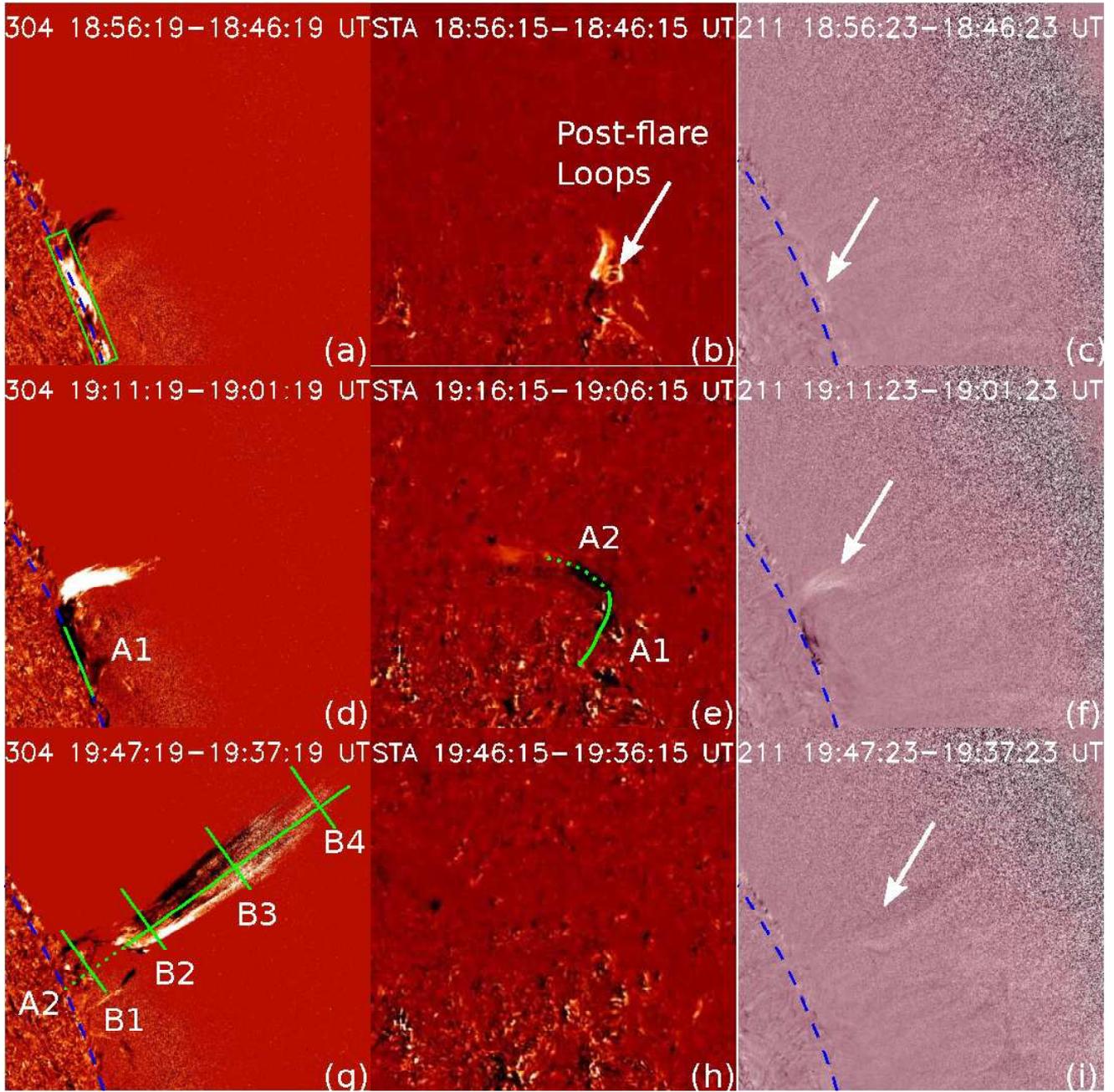}
\caption{Left column: Difference images taken by SDO/AIA at 304\AA\
passband. The FOV of the images is $430''\times430''$. Middle column:
Difference images from STEREO-A/EUVI at the same passband. The FOV
is $450''\times450''$. Since STEREO-A was 120$^\circ$ apart away
from SDO on 2012 July 8, the SDO limb event right happened ondisk in
the view of STEREO-A. Right column: Difference images taken by SDO/AIA at 211\AA\
passband with the same FOV of the images in the left column. The white arrow in the middle column
denotes the post-flare loops and those in the right column mark the hot-component of the jet.} \label{ov}
\end{center}
\end{figure*}

This fact is closely related with another common phenomenon,
apparent rotational/torsional motion of jet plasma during its
ascending and/or descending phase \citep[e.g.,][]{Xu1984,
Shibata1985, Canfield1996, Jibben2004, Shimojo2007, Liu2009}. A
well-accepted picture is that the reconnection between twisted loops
and untwisted open field lines causes helicity transferred from
loops to open field lines and therefore makes plasma moving upward
helically along the path through nonlinear torsional Alfv\'en waves
\citep[e.g.,][]{Pariat2009} or Lorentz force working
\citep{Shibata1985}. It is interesting to see which one is
more appropriate, or if there is alternative explanation.

One may find that the additional force pushing a jet unusually high
is probably just the one driving the apparent rotational motion. In
many cases, the jet keeps rising after reconnection. It implies
that, during a jet, the magnetic free energy is released through two
different ways. One is reconnection, and the other is
post-reconnection relaxation of magnetic field structure. Related to
the issue raised for the rotational motion, an interesting question
is how and how significantly the latter contributes to the jet
kinetic energy, or in other words, when and how a jet gains its
kinetic energy.

Here, we will try to address this issue by investigating a
prominence-like jet that was observed by SDO/AIA \citep{Lemen2012}
and STEREO/SECCHI EUVI \citep{Howard2008} simultaneously. Thanks to
the high-resolution, high-cadence, multi-wavelength and multi-point
observations from SDO and STEREO, we are for the first time able to
accurately assess its energy budget in observations.

\section{Overview}

The event located off the north-west limb of the Sun, a bit north to
the active region (AR) 11513. Two successive jets
can be found at the same place from 18:00 to 21:00 UT on 2012 July 8
in various EUV passbands (see multi-wavelength movie M1). They were
the most visible in 304{\AA}, and also showed weak signatures in
the hotter channel 211\AA\ (as seen in Fig.\ref{ov}). But the jets were hard to be seen in emission lines
with temperature higher than 211\AA\, suggesting that they are cool-component-dominant jets
with temperature generally below 2 MK. An online movie M2
generated from AIA 304\AA\ passband shows the detailed ejection
process of the two jets. The first jet was a minor one with a life
time of about one hour. It began to ascend at about 18:00 UT,
reached its maximum height of about $90$ Mm 35 minutes later, and
then fell back to the solar surface at about 18:56 UT. The second
jet is much more significant, which took place right after the first
one and lasted for about 2 hours. In this study we will focus on the
second jet.

\begin{figure}[tbh]
\centering
\includegraphics[width=\hsize]{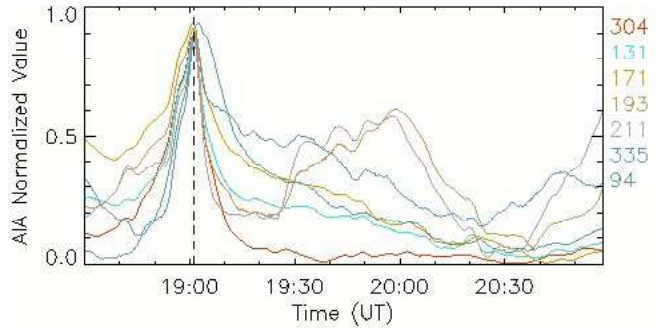}
\caption{Normalized light curve derived from the integral emission
from the brightening region (indicated by the green box in
Fig.~\ref{ov}a).} \label{cur}
\end{figure}

The second jet was triggered by a micro-flare, which caused obvious
enhancements of the EUV emissions at various wavelengths as shown in
Figure~\ref{cur} with the peak at about 19:01 UT (indicated by the
black dashed line). The core of the micro-flare manifesting as a
brightening point first appeared around 18:48 UT at the latitude of
about 22$^\circ$, and then moved on the solar surface to the
latitude of about $25^\circ$, which probably suggests that the
reconnection point was moving. Meanwhile, several brightening small
loops appeared beside the brightening point. Accordingly some
prominence-like materials traveled along a tunnel lying on the solar
surface between the latitude of 22$^\circ$ and $25^\circ$ at the
beginning (see Fig.~\ref{ov}a and \ref{ov}b).

These materials formed the jet, which started to rise straightly at
about 19:05 UT, slightly away from the local radial direction.
According to Figure~\ref{cur} and the online movie M2, the
micro-flare faded away around 19:11 UT, suggesting that the
reconnection probably lasted for about 23 minutes. At that time the
jet was confined within a tunnel with a width of about 15 Mm
(Fig.~\ref{ov}d). The rising of the jet could be found in the
STEREO-A/EUVI images (Fig.~\ref{ov}e), but the signature is weak due
to the relatively low resolution and low cadence of STEREO data.

The jet kept rising after 19:11 UT. It quickly expanded to about 35
Mm wide in a short distance, and gradually grew to about 50 Mm wide
when it reached its maximum height of about 292 Mm at about 19:47 UT
(Fig.~\ref{ov}g). After then, the jet began to fall back. During the
whole ejection process, we can find continuous rotational motion
around the jet axis. From the AIA 304\AA\ movie, one can clearly
distinguish many pieces of prominence-like materials rotating like a
rigid object. In lots of previous studies, the rotational motion
appeared only in the ascending phase. Thus torsional Alfv\'en waves
could be a driver of it. However, in this case, the jet plasma kept
rotated during its descending phase, and the rotational period did
not change significantly as will be seen in Sec.\ref{sec_rotation}.
This is hard to be explained only by a upward-propagating wave train.
This fact spurs us to figure out the real physics behind it. Is the
rotational motion the manifestation of real motion of plasma along a
twisted magnetic field lines, or a rigid rotation of a bundle of
untwisted magnetic field lines? To solve this puzzle, we analyze the
axial motion and rotational motion, respectively, in the next two
sections.

\section{Axial Motion}

\begin{figure*}
\begin{center}
  \includegraphics[width=\hsize]{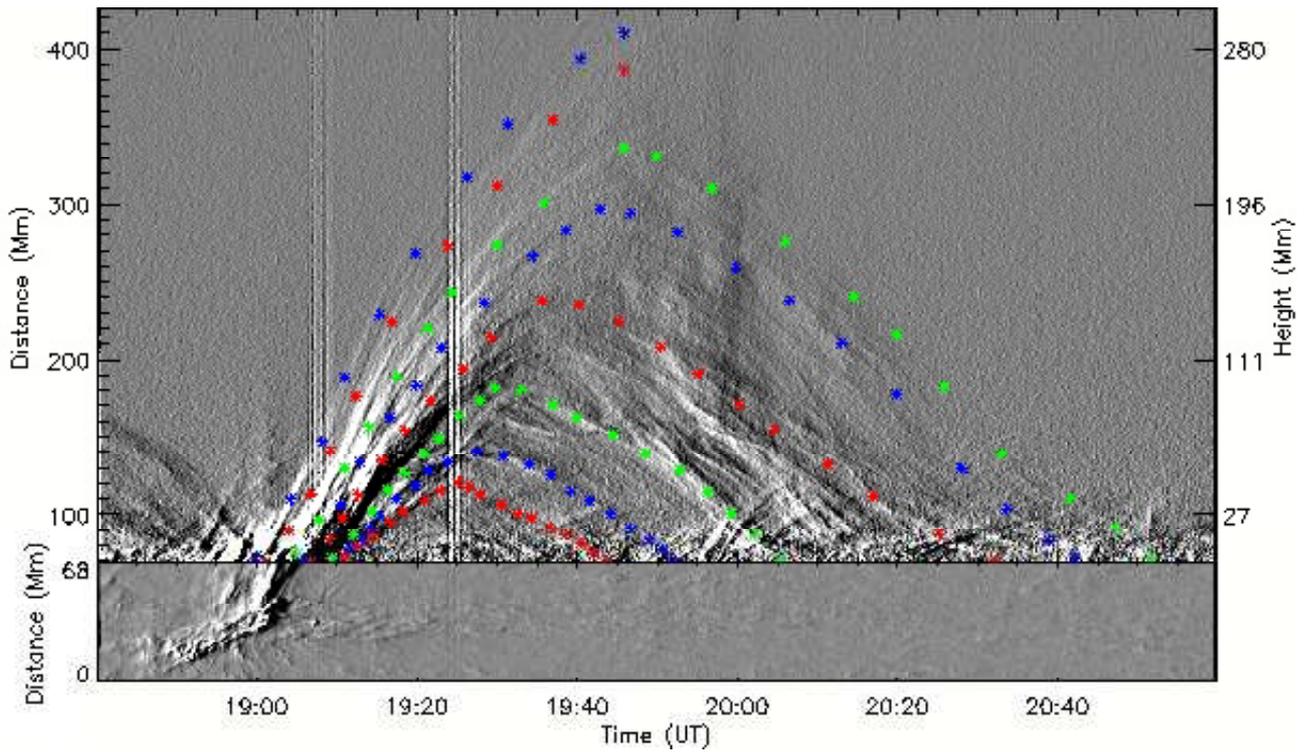}
  \caption{Lower part: De-projected running-difference space-time plot generated from slice A1 (see Fig.\ref{ov}d).
  Upper part: De-projected running-difference space-time plot generated from slice A2 (see Fig.~\ref{ov}g).
  The left vertical axis gives the distance from the start point along the slice, and the corresponding height
  from solar surface is marked on the right vertical axis.}\label{ax}
  \end{center}
\end{figure*}

\begin{figure*}
\begin{center}
\includegraphics[width=\hsize]{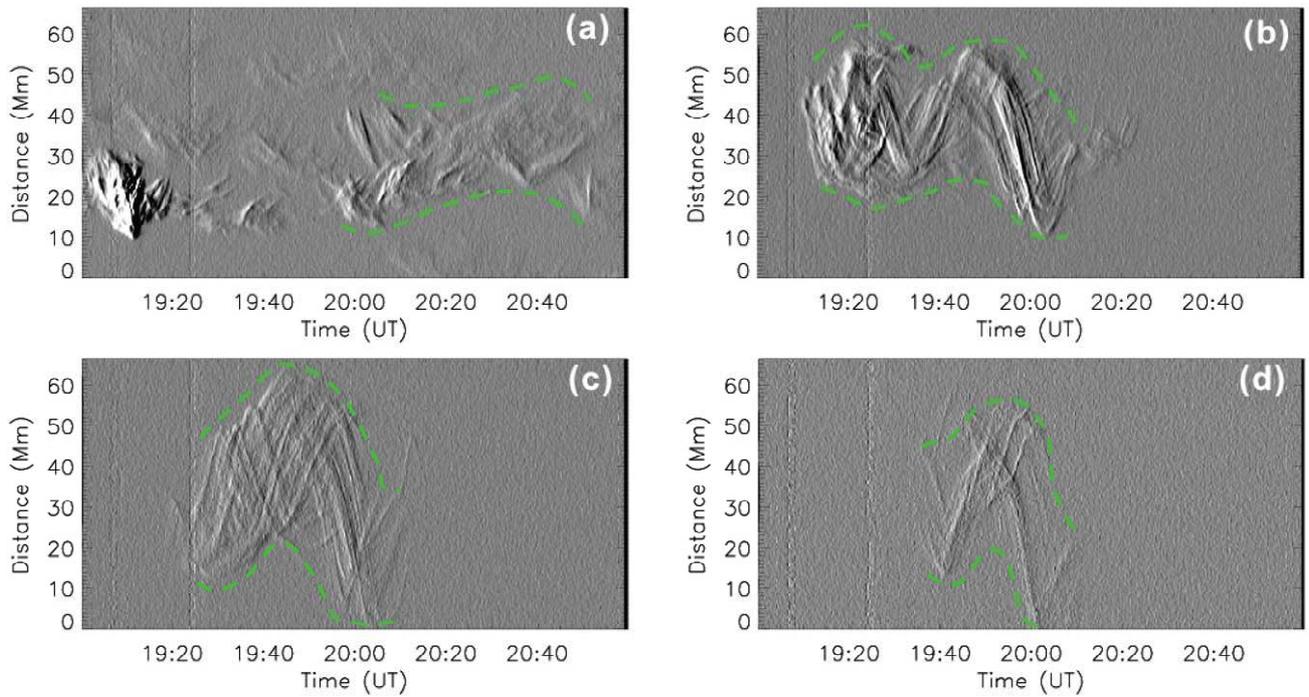}
\caption{Running-difference space-time plot generated from slice B1,
B2, B3 and B4, as marked in Figure~\ref{ov}g. These four slices are
all perpendicular to the jet's axis. Dashed lines indicate the
boundary of the wriggling jet tunnel.} \label{tor}
\end{center}
\end{figure*}

To study the axial motion of the jet, a slice is placed along the
jet tunnel. The slice has two segments; one (labeled as A1 in
Fig.\ref{ov}d) lies on the surface and the other (A2, Fig.\ref{ov}g)
stands upward straightly. The segment A1 was visible for both SDO
and STEREO-A (solid green line in Fig.\ref{ov}e), the projection
effect could be easily removed. For segment A2, it is visible for
SDO but not all for STEREO-A; only the lower part of A2 can be
recognized in STEREO-A/EUVI images (as indicated by the dotted green
line in Fig.\ref{ov}e). Thus, we assume that segment A2 is straight
and use its lower part to correct the projection effect of A2. It is
derived that the segment is about 30$^\circ$ away from the
plane-of-sky in the view of SDO. A space-time plot generated from
the slice is shown in Figure~\ref{ax}, in which the projection
effect has been corrected.

An obvious acceleration could be seen in the plot when the jet moved
on the surface. A quadratic fitting to the tracks in the low part of
Figure~\ref{ax} suggests that the acceleration is about 300 m
s$^{-2}$. The jet moved with an average velocity of about 95 km
s$^{-1}$ and then turned upward with a speed of about 160 km
s$^{-1}$ overall. When the jet moved upward, we may distinguish many
small sub-jets in it, which are shown as bright-dark alternating
stripes in the upper part of Figure~\ref{ax}. These sub-jets were
expelled successively. They experienced acceleration at the
beginning, and then turned to deceleration. We tracked eight
sub-jets as indicated by color-coded asterisks in Figure~\ref{ax}.
The initial speeds of these sub-jets ranged from about 57 to 170 km
s$^{-1}$, and through an acceleration, they reached maximum speeds
in a range of about 79 to 238 km s$^{-1}$ around 19:11 UT, when the
brightening faded away. The earlier sub-jet has a larger
acceleration and larger speed. These results are consistent with the
fact that the micro-flare decayed with time.

After reaching the maximum upward speed, these sub-jets began to
decelerate. Overall, the sub-jets were decelerated during the whole
ascending phase, and the deceleration ranged from about $-21$ to
$-67$ m s$^{-2}$. These values are smaller than the local gravity,
even if the uncertainty (see the note in Table~\ref{tb0}) is taken
into account. It means that continuous upward force exists after the
reconnection. These sub-jets finally reached up to a height from
about 42 to 292 Mm (or 120 to 410 Mm in distance along the jet
tunnel). Consistent with the speeds obtained before, the earlier
sub-jets experienced a longer ascending phase and reached a higher
height, which is clearly shown in Figure~\ref{ax}.

The descending speeds of these sub-jets were about $-44$ to $-70$ km
s$^{-1}$ with an average downward acceleration from $-3$ to $-32$ m
s$^{-2}$, which were all smaller than those during the ascending
phase. A direct consequence is that the duration of the descending
phase is obviously longer than that of the ascending phase.
Table~\ref{tb0} lists the kinematic parameters for the eight
selected sub-jets.

\section{Rotational Motion}\label{sec_rotation}

In order to analyze the rotational motion of the jet, we place four slices
perpendicular to the jet tunnel at the height of 30, 90, 180 and 270 Mm,
respectively (marked by B1 to B4 in Fig.~\ref{ov}g). Figure~\ref{tor}
shows the space-time plots generated from the four slices, in which the
end of a slice at the higher latitude is referred as zero and stripes with
positive slopes indicate motion of material from higher latitude toward
lower latitude.

From these plots, we can see many sine-like tracks, suggesting
rotational motion in the jet tunnel. Such sine-like tracks appeared
during both ascending and descending phases. In particular, these
rotating materials seemingly concentrated near the surface of the
jet tunnel. According to these tracks, we find that the jet tunnel
was wriggling slightly, as indicated by the dashed lines. It is
estimated that the width of the jet tunnel at the four heights is
about 12, 30, 40 and 45 Mm, respectively.

Assuming that the jet tunnel is a cylinder with a varying radius,
the real rotational speed could be derived by fitting these curves
with a sine function. Table~\ref{tb2} gives the derived parameters
for the rotational motion. It is found that their period is around
1270 s, and there is no significant difference in the period at
different heights. The real rotational speed at the four different
heights is therefore about 32, 74, 97 and 106 km s$^{-1}$,
respectively. Although these results suffer from a large error, they
suggest that the jet material rotated faster and faster as it
ascended and then slowed down when it fell back.

As mentioned before, for such an apparent rotational motion, there
could be two different interpretations. One is that prominence-like
materials move along twisted or helical magnetic field lines, and
the other is that a bundle of straight magnetic field lines rotate
as a rigid body in which prominence-like materials move up and down.
If the first interpretation is the case, we expect that the turns
the materials rotated around the jet tunnel within a given distance
should be the same for different sub-jets. For sub-jet 1 (see
Table~\ref{tb0}), the turns per unit length is
$n=\frac{1}{T_cv_{avg}}\approx\frac{1}{1270\textrm{(s)}\times126\textrm{(km
s}^{-1}\textrm{)}}=0.006$ Mm$^{-1}$. Using this number to constrain
sub-jet 7 or 8, we may derive that the expected period of them
should be about 2381 or 2976 s, which is much larger than the
observed period given in Table~\ref{tb2} even if the uncertainty is
taken into account, and cannot be found in the space-time plots.

Thus, the second interpretation is more appropriate. In this
scenario, the jet tunnel above the limb consists of
straight/untwisted open magnetic field lines. They rotated due to
the reconnection at the jet root, which connected the untwisted open
magnetic field lines to a bunch of highly twisted magnetic field
loops and caused the helicity transported from the twisted fields
into untwisted fields. The brightening and small loops shown in the
first and second panels of Figure~\ref{ov} are the signatures. The
transport process therefore manifested a rotational motion.

\begin{table*}
\linespread{1.5} \caption{Kinematic parameters of eight sub-jets in
the axial direction.}
\begin{center}
\begin{tabular}{c|ccccc|ccc|c}
\hline
& \multicolumn{5}{c|}{Ascending Phase} & \multicolumn{3}{c|}{Desceding Phase} & \\
Sub-jet& $v_{ini}$  & $v_{max}$ & $v_{avg}$ & $a_{avg}$ & $T_a$ & $v_{avg}$ & $a_{avg}$ & $T_d$ & $H_{max}$ \\
\hline
1&170$\pm$10 & 238$\pm$16 & 126$\pm$2 & -56$\pm$5  & 2754 & --        & --        &   -- & 292 \\
2&144$\pm$14 & 193$\pm$21 & 124$\pm$2 & -49$\pm$5  & 2693 & --        & --        &   -- & 272 \\
3&150$\pm$12 & 178$\pm$19 & 110$\pm$2 & -56$\pm$6  & 2468 & -70$\pm$1 &  -6$\pm$2  & 3943 & 229 \\
4&140$\pm$10 & 174$\pm$18 & 102$\pm$2 & -48$\pm$7  & 2222 & -68$\pm$1 & -17$\pm$3  & 3564 & 196 \\
5&125$\pm$19 & 135$\pm$33 & 100$\pm$3 & -46$\pm$12 & 1730 & -53$\pm$1 &  -3$\pm$3  & 3340 & 145 \\
6&104$\pm$16 & 122$\pm$30 &  90$\pm$4 & -35$\pm$23 & 1229 & -52$\pm$2 & -28$\pm$8  & 2130 & 98 \\
7&75$\pm$24  & 126$\pm$53 &  70$\pm$5 & -67$\pm$31 & 1075 & -49$\pm$3 & -32$\pm$15 & 1495 &  61 \\
8&57$\pm$27  &  79$\pm$50 &  56$\pm$5 & -21$\pm$40 &  957 & -44$\pm$4 & -10$\pm$27 &  1106 & 42 \\
\hline
\end{tabular}\\
$v_{ini}$, $v_{max}$ and $v_{avg}$ are the initial, maximum and
average speed, respectively, in units of km s$^{-1}$. $a_{avg}$ is
the average acceleration in units of m s$^{-2}$. $T_a$ and $T_d$ are
the duration in units of second. $H_{max}$ is the maximum height a
sub-jet reached, which is units of Mm. The uncertainty in the
velocity and acceleration is estimated through the fitting procedure
by assuming a 10-pixel error in measuring height (corresponding to a
5-Mm error in distance). Positive values correspond to the upward
direction.
\end{center}
\label{tb0}
\end{table*}

\section{Energy budget}

During the jet process, some prominence-like materials reached as
high as 290 Mm or so, suggesting a significant release of magnetic
energy. The release process of the magnetic energy obviously has two
stages. The first stage is from 18:48 to 19:11 UT. During the stage,
a micro-flare took place and then faded away, and meanwhile, the jet
traveled on the solar surface and then climbed up to as high as 100 Mm.
The second stage is from 19:11 UT to the end of the event. During the
stage, the jet continuously ascended until about 19:47 UT and then
fell back. The acceleration is significantly smaller than the solar
gravity.

For most of such events, the magnetic energy was released through
two ways. One is the magnetic reconnection, during which the free
magnetic energy is directly converted to produce both thermal and
non-thermal emissions and kinetic energy of plasma jets. The
resultant magnetic structure through the reconnection may not be at
a stable state. It will further relax its configuration to lower
energy level. This becomes the other way to release the free energy.
For the first stage, both the ways may take effects, and for the
second stage, the second way is the only one. It is not new for us
that the magnetic energy could be released in such ways, but it is
really unclear whether only one of them or both are important for
the ejecta. Flare is much easier to catch people's eyes and usually
thought to be the major approach to convert magnetic energy into
plasma kinetic energy. How much magnetic energy will be further
released after a flare? This question is now be addressed below.

\begin{figure}
\begin{center}
\includegraphics[width=\hsize]{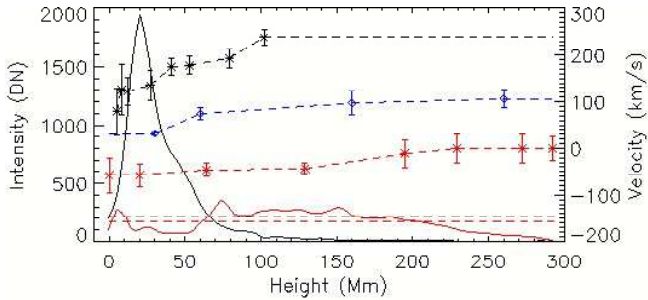}
\caption{Solid lines show the integrated intensity of EUV 304\AA\
emission over the cross-section of the jet as a function of height,
in which the background emission is subtracted. The black line is
calculated at 19:11 UT and the red line at 19:47 UT. The mean
intensity at the two instants is indicated by two horizontal dashed
lines, which are very close to each other. Asterisks connected with
dashed lines indicate the axial velocities of the eight sub-jets,
and the diamonds connected with dashed lines are the rotational
velocities at the four heights.} \label{den}
\end{center}
\end{figure}

Here we compare two instants. One is at 19:11 UT when the
micro-flare ended and the jet roughly reached a maximum ascending
speed (Fig.\ref{ov}d), and the other is at 19:47 UT when the jet
reached the maximum height (Fig.~\ref{ov}g). Figure~\ref{den} shows
the emission intensity, $I$, as a function of height at the two
instants. The emission intensity is calculated based on images in
EUV 304\AA\ passband, and it is an integrated value over the
cross-section of the jet cylinder at any given height. Here the
background emission is removed by subtracting the average value of
the pixels surrounding the jet.

\begin{table}
\linespread{1.5} \caption{Kinematic parameters of the rotational
motion of the jet.}
\begin{center}
\begin{tabular}{c|cccc}
\hline
&$H$ & $D$  & $T_c$  & $v_{\phi}$ \\
\hline
 B1 & 30 & 12 & 1180$\pm$120 & 32$\pm$3\\
 B2 & 90 & 30 & 1270$\pm$230 & 74$\pm$13 \\
 B3 & 180 &40 & 1290$\pm$330 & 97$\pm$25\\
 B4 & 270 &45 & 1330$\pm$250 & 106$\pm$20 \\
\hline
\end{tabular}
\end{center}
$H$ is the height of the four slices in units of Mm, $D$ is the width (assuming being diameter) of the jet tunnel in units of Mm,
$T_c$ is the period of the rotational motion in units of seconds,
and $v_\phi$ is the rotational speed in units of km s$^{-1}$.
\label{tb2}
\end{table}

The average value $I_0$ of the intensity of the whole jet is about
216 and 168, respectively, in units of digital number (DN) at the
two instants (as indicated by the two horizontal dashed lines in
Fig.~\ref{den}). For prominences observed in 304\AA\ emission line,
which is optically thick, it could be accepted that $\rho\propto I$,
where $\rho$ is the density. Thus, the product of the average
intensity and the height could be a proxy of the mass of the jet
material. The difference of the average intensity between the two
instants suggests that the mass is not the same, but the difference
is relatively small. It may be caused by the errors in measurements
or the shielding effect in the optically thick medium.

In order to make the two instants comparative, we investigate the
energy per unit mass. According to the distribution of the intensity
shown in Figure~\ref{den}, the gravitational potential energy per
unit mass gained by the jet can be calculated by
$E_{g}=GM_\odot\frac{\int_{0}^{H}I\left(\frac{1}{R_\odot}-\frac{1}{h+R_\odot}\right)dh}{\int_{0}^{H}Idh}$.
The kinetic energy per unit mass of the jet consists of two
components. One is the linear kinetic energy and the other is the
angular kinetic energy, which are given by $E_{l}=\frac{\int_{0}^HI
v^2dh}{2\int_{0}^HIdh}$ and $E_a=\frac{\int_{0}^HI
v_\phi^2dh}{2\int_{0}^HIdh}$. The linear velocity could be read from
Figure~\ref{ax} and the angular velocity from Table~\ref{tb2}. These
velocities have been marked as symbols in Figure~\ref{den}. The
velocity between the symbols is obtained by using linear
extrapolation, and the velocity outside the symbols just chooses the
value of the nearest symbol (as indicated by the dashed lines
connecting the symbols).

\begin{table}
\linespread{1.5} \caption{Energies of the jet at two instants.}
\begin{center}
\begin{tabular}{c|cccc}
\hline
Time & $E_g$ & $E_{l}$  & $E_a$ & $E_t$ \\
\hline
 19:11 UT & 0.88 & $1.15^{+0.44}_{-0.35}$ & $0.11^{+0.03}_{-0.04}$ & $2.14^{+0.47}_{-0.39}$ \\
 19:47 UT & 2.83 & $0.08^{+0.08}_{-0.04}$ & $0.37^{+0.18}_{-0.14}$ & $3.28^{+0.26}_{-0.18}$\\
\hline
\end{tabular}\\
Energies are in units of $10^{10}$ J kg$^{-1}$.
\end{center}
\label{tb3}
\end{table}

Table~\ref{tb3} lists the energies per unit mass. First of all, the
total energy at 19:47 UT is larger than that at 19:11 UT. Their
difference $\Delta E$ is about $1.14\times10^{10}$ J kg$^{-1}$,
which is about 3 times of the uncertainties of total energy at each
instant, suggesting a significant difference. The micro-flare ended
at 19:11 UT, which means that there was a continuous conversion from
magnetic energy to potential and kinetic energies after the
reconnection. The observed rotational motion suggests that an
untwisting process at the root of the jet is responsible for the
energy conversion, through which the post-reconnection magnetic
field structure relaxes to a lower energy state. The amount of the
released magnetic energy could be alternatively estimated from the
measurements of the accelerations of sub-jets. Since their
acceleration (see $a_{avg}$ listed in Table~\ref{tb0}) is much
smaller than the gravity, there must be additional force
$f=m(a_{avg}-g)$ acting on the jet, where $m$ is the mass and
$g=-274$ m s$^{-2}$ is the gravity. The work per unit mass done by
the force is $W=fH/m=(a_{avg}-g)H$. According to the values of
$a_{avg}$ and $H$ given in Table~\ref{tb0}, it is easily inferred
that $W$ is on the order of $10^{10}$ J kg$^{-1}$, which is
consistent with $\Delta E$ derived above.

Usually reconnection produces straight plasma beams, like a jet.
Thus it is reasonable to assume that the angular kinetic energy
$E_a$ should mostly come from the untwisting process. We may infer
that, for the kinetic energy (including the potential energy) of the
jet, the contribution from the reconnection, i.e., the micro-flare,
is $E_{r}=E_{t0}-E_{a0}=2.03\times10^{10}$ J kg$^{-1}$, and the
contribution from the untwisting process in the ascending phase is
$E_{ua}=E_{t1}-E_{t0}+E_{a0}=1.25\times10^{10}$ J kg$^{-1}$, where
the subscript $0$ and $1$ refers to the instant of 19:11 and 19:47
UT, respectively. Moreover, by considering that (1) the rotational
motion continuously existed during the descending phase, (2) the
rotational velocity and period are almost as the same as those
during the ascending phase and (3) the duration of the descending
phase is about 1.57 times of that of the ascending phase (ref.
Table~\ref{tb0}), we derive that the contribution of the untwisting
process during the whole event is roughly
$E_u=2.57E_{ua}=3.2\times10^{10}$ J kg$^{-1}$, which is 1.6 times of
the kinetic energy that could be injected by the reconnection. Even
if considering the kinetic energy produced by a reconnection/flare
is only a small fraction (about 10\%) of its total released energy
\citep[e.g.,][]{Woods2004, Benz2008,
Reeves2010, Emslie2012}, the contribution of the
untwisting process is still significant, which is about 16\% of the
total released magnetic free energy by a reconnection.

\section{Conclusions and Discussions}
We presented the observational features of a prominence-like jet
event observed by SDO/AIA and STEREO/SECCHI EUVI simultaneously on
2012 July 8. Like most jets observed before, it was triggered by a
micro-flare accompanied with several small brightening loops,
suggesting a weak reconnection. After obtaining initial momentum,
the jet traveled along a tunnel to reach a height of about 292 Mm
above the solar surface, and then returned back to the Sun. During
the whole process, the acceleration in radial direction is
significantly smaller than solar gravity, implying an additional
force acting on the jet plasma even after the reconnection.
All these observations fit well the classical jet
model as proposed in Figure 4 of the paper by \citet{Shibata1996}.


The magnetic free energy is released through two ways during the
jet. One is reconnection and the other is the magnetic field
relaxation after the reconnection. By analyzing its motion and
energy budget, we find that the magnetic field relaxation after the
reconnection makes a significant contribution for the jet to gain
kinetic energy, which is about 1.6 times of the contribution made by
reconnection, and about 16\% of total magnetic free energy that
could be released by the reconnection.

Rotational motion is a manifestation of continuously conversion of
magnetic energy into kinetic energy through a way rather than the
reconnection. In this case, we believe that the twisted loops, which
are connected to the untwisted magnetic field lines, drive the
rotation. But different from traditional picture, the rotation is
probably not mainly caused by torsional Alfv\'en wave
\citep{Pariat2009}. The reason is that (1) the rotation appeared in
both ascending and descending phases, and (2) the additional force
preventing the jet plasma falling back is even larger during the
descending phase, which caused the descending phase much longer than
the ascending phase. These new findings are not expected by the
classical jet model. What is the physics behind them is worthy
of further study.

A similar picture showing the rotation of a bundle of untwisted (or
weak twisted) magnetic field lines could be found in a recent study
of solar tornadoes/cyclones \citep{Wedemeyer-Bohm2012}. In
their case, vortex flows at base rather than reconnection drive up
flow. Solar cyclones are found to be a ubiquitous phenomenon in the
solar atmosphere \citep[e.g.,][]{Brandt1988,
Wedemeyer-Bohm2009, Attie2009,
Zhang2011, Wedemeyer-Bohm2012, Li2012,
LiuJJ2012, Su2012}. Jets also a ubiquitous phenomenon in the
solar atmosphere. Thus we may conjecture that the rotational motion
generated during jets, which is usually observed from side-view, and
the cyclones that are usually observed from top-view may be probably
the same thing, or at least jets are a subset of solar cyclones. For
this case, we are unable to make deeper analysis on this issue,
because of the low resolution and low cadence of STEREO-A data,
although STEREO-A observed the event from another angle of view.

In this case, the conversion rate per unit mass of the magnetic
energy to kinetic energy is about $E_u/(T_a+T_d)=5\times10^6$ J
kg$^{-1}$ s$^{-1}$. By assuming a typical number density of about
10$^{10}$ cm$^{-3}$ of the jet plasma \citep{Roy1973}, the
conversion rate per unit volume is about $8\times10^{-5}$ J m$^{-3}$
s$^{-1}$, and the momentum flux is about $1.7\times10^{4}$ J
m$^{-2}$ s$^{-1}$ by considering a length scale of about 200 Mm. The
radiation of hot corona requires a energy flux of about
$3\times10^2$ J m$^{-2}$ s$^{-1}$ into thermal energy
\citep[e.g.,][]{Withbroe1977, Aschwanden2006}, which is 2\% of the
momentum flux of the jet. In other words, the local corona could be
heated as long as only a very small fraction of kinetic energy
carried by the jet is dissipated. Thus, we believe that jets are able
to heat local corona when they get kinetic energy, as suggested in
many previous studies for spicules and X-ray jets
\citep[e.g.,][]{Tsiropoula2004, DePontieu2007, Shibata2007,
Cirtain2007}.

\acknowledgments{We acknowledge the use of data from AIA instrument
onboard Solar Dynamics Observatory (SDO) and EUVI instrument onboard Solar TErrestrial RElations Observatory (STEREO).
This work is supported by grants from the CAS (the Key Research Program KZZD-EW-01-4,
100-talent program, KZCX2-YW-QN511 and startup fund), 973 key
project (2011CB811403), NSFC (41131065, 40904046, 40874075, and
41121003), MOEC (20113402110001) and the fundamental research funds
for the central universities.}

\bibliographystyle{agufull}

\end{document}